# Comparison of Spin-Wave Modes in Connected and Disconnected Artificial Spin Ice Nanostructures Using Brillouin Light Scattering Spectroscopy

Avinash Kumar Chaurasiya,[1] Amrit Kumar Mondal,[1] Jack C. Gartside,[2] Kilian D. Stenning,[2] Alex Vanstone,[2] Saswati Barman,[3] Will R. Branford,[2,4] and Anjan Barman[1,*]

[1]Department of Condensed Matter Physics and Material Sciences, S. N. Bose National Centre for Basic Sciences, Block – JD, Sector-III, Salt Lake, Kolkata 700 106, India

[2]Blackett Laboratory, Department of Physics, Imperial College London, SW7 2AZ, United Kingdom

[3]Institute of Engineering and Management, Sector-V, Salt Lake, Kolkata 700 091, India

[4]London Centre for Nanotechnology, Imperial College London, SW7 2AZ, United Kingdom

**ABSTRACT:** Artificial spin ice systems have seen burgeoning interest due to their intriguing physics and potential applications in reprogrammable memory, logic and magnonics. Integration of artificial spin ice with functional magnonics is a relatively recent research direction, with a host of promising results. As the field progresses, direct in-depth comparisons of distinct artificial spin systems are crucial to advancing the field. While studies have investigated the effects of different lattice geometries, little comparison exists between systems comprising continuously connected nanostructures, where spin-waves propagate *via* dipole-exchange interaction, and systems with nanobars disconnected at vertices where spin-wave propagation occurs *via* stray dipolar-field. Gaining understanding of how these very different coupling methods affect both spin-wave dynamics and magnetic reversal is key for the field to progress and provides crucial system-design information including for future systems containing combinations of connected and disconnected elements. Here, we study the magnonic response of two kagome spin ices *via* Brillouin light scattering, a continuously connected system and a disconnected system with vertex gaps. We observe distinct high-frequency dynamics and magnetization reversal regimes between the systems, with key distinctions in spin-wave localization and mode quantization, microstate-trajectory during reversal and internal field-profiles. These observations are pertinent for the fundamental understanding of artificial spin systems and broader design and engineering of reconfigurable functional magnonic crystals.

Artificial spin ice (ASI) systems are engineered magnetic metamaterials designed by nanopatterning strongly interacting single domain nanoislands in a geometrically frustrated array.[1] Frustrations in physical systems emerge from inability to simultaneously minimize all interactions.[2] In 2006, Wang *et al.* carried out pioneering experiments on lithographically defined arrays of interacting nanomagnets and explored interesting physics analogous to spin ice materials (Pyrochlores, $Dy_2Ti_2O_7$ *etc.*).[3] This triggered research efforts to investigate ASI for exotic fundamental physics as well as potential applications such as reprogrammable memory, logic and more recently reconfigurable magnonics.[4-8] Studies on ASI have revealed the occurrence of classical spin liquid states,[9] Coulomb phases,[10] monopole-like excitation[11,12] and spin fragmentation.[13] In the context of experimental probing of the spin dynamics, ferromagnetic resonance (FMR) technique has been mostly employed which is based on the global excitation and probing of a large area ASI subject to an externally swept field.[14-24] Because of the large array of nano islands in such studies, it is quite unclear to attribute that the observed changes in FMR frequencies rely upon the number of nanobars that have been reversed or whether the precise spin configuration of the array plays a crucial role. Understanding spin dynamics in a limited array size gives a detailed understanding of the local switching and spin-wave (SW) behaviour of ASI and provides a good alternative with which to compare results of the widely used FMR technique. The main focus of this study is to directly compare continuously-connected arrays where elements touch at vertices with disconnected, vertex-gapped arrays. While studies exist on both connected and disconnected arrays, they have examined them in isolation looking at a single regime per study with different lattice geometries, dimensions and different experimental techniques precluding direct side-by-side comparison. This hinders researchers attempting to learn from the literature how introducing continuously-connected, dipole-exchange mediated interactions to their system will differ from vertex-gapped dipolar field interactions, an issue which this study aims to address.

Here, we have performed a comparative study of SW dynamics and evolution through magnetic field driven microstate transitions in connected and disconnected kagome ASI (c-ASI and d-ASI hereafter) made of $Ni_{80}Fe_{20}$ (permalloy, Py hereafter). We study magnetically saturated and disordered states *via* Brillouin light scattering (BLS) and micromagnetic simulation. Key findings are distinct differences between SW-mode localization with modes well-localized in specific array elements for the dipolar-coupled, disconnected case and a combination of low-frequency, spatially localized and higher-frequency spatially-dispersed modes in the dipole-exchange mediated connected system. This has implication for future nanomagnetic system design, both ASI-based and for wider family of magnonic crystals and offers insight into how combinations of both disconnected and connected elements may be of future physical interest and technological benefit.

**RESULTS AND DISCUSSION**

$90 \times 90$ µm$^2$ array of 25-nm-thick Py kagome ASI were fabricated using electron-beam lithography (EBL) and lift-off process. The samples were then capped with 3.7 nm of $Al_2O_3$ to protect it from oxidation. For the c-ASI sample, the nanobars have length (*l*) and width (*w*) of 300 nm and 80 nm, respectively. The d-ASI has *l* and *w* matching the c-ASI but a 50 nm vertex gap (defined as the distance between nanobar-end and vertex centre) is introduced.

Figure 1a-b represent the scanning electron microscopy (SEM) images for connected (c-ASI) and disconnected (d-ASI) kagome ASI samples, respectively. The variation in the measured dimensions is: ±3% in length, ±2% in width and ±1.5% in the inter-bar distance for d-ASI, while the heights of the nanobars are found to vary by ±3% measured *via* atomic force microscope (AFM) (see Supporting Information, Figure S1). The SEM images confirm good quality of the samples. Figure 1c-d show the remanent state MFM images for the samples c-ASI and d-ASI, respectively. The MFM images reveal that the magnetization orientation for both c-ASI and d-ASI obey the spin ice rule, *i.e.,* "two-in one-out" or "one-in two-out" with no observed vertices exhibiting three like-polarity magnetic charges. MFM images simulated at remanence using LLG simulator for both c-ASI and d-ASI (Figure 1e-f) show excellent agreement with the experimental MFM images. Figure 1g shows the schematic of the BLS measurement geometry.

**Magnetization Reversal of Kagome ASI**

Figure 2a represents the magnetic hysteresis loops measured by longitudinal magneto-optical Kerr effect (MOKE) for both samples c-ASI and d-ASI. The arrays exhibit coercive fields of 370 Oe for c-ASI and 450 Oe for d-ASI, remanence of 70% for c-ASI and 82% for d-ASI, and a two-step reversal for c-ASI as opposed to a single step reversal for d-ASI. The remanence is less than 100% because there is a bending of the magnetization away from the bar axes in field; there are no bar reversals at remanence. The difference in structure of c-ASI and d-ASI arrays leads to a lower coercivity for the c-ASI structure as once domain walls (DWs) nucleate at the array edges they may propagate through the connected vertices, whereas the d-ASI sample requires a separate DW nucleation event for every nanobar and corresponding higher coercivity. This two-step feature occurs because the DW propagation 1D avalanches can only switch approximately 50% of the diagonal bars, the remainder have to reverse by intra-array DW nucleation at higher field.[25] Figure 2b shows the simulated hysteresis loops which are in qualitative agreement with the experimental loops. We will further investigate how the magnetization reversal behaviours affect the high frequency SW dynamics of these samples.

**Spin-Wave Dynamics of Kagome ASI**

Figure 3a,c show the representative BLS spectra for c-ASI and d-ASI taken at $k \approx 0$ in the Damon-Eshbach (DE) geometry at different $H$ values. The corresponding simulated SW spectra are shown in Figure 3b,d.

In the BLS experiment, $H$ was varied in the range +1.4 kOe $\leq H \leq$ -1.4 kOe, tracing the upper branch of the magnetic hysteresis, and SW frequencies were recorded from BLS spectra. For both c-ASI and d-ASI, we observe three dominant modes whose frequencies and BLS intensities vary appreciably and non-monotonically with $H$. In addition, a stark mode frequency difference is observed between c-ASI and d-ASI due to their differing inter-nanobar coupling mechanism as also evidenced in the MOKE hysteresis loops. Figure 3e-f are the experimental BLS spectra taken at $H$ = -415 Oe and -550 Oe for c-ASI and d-ASI, respectively, halfway through magnetization reversal as measured *via* MOKE. It is clear from this figure that two additional higher frequency SW modes (M4 and M5) appear for both samples. These modes grow at the expense of modes M2 and M3 as the magnitude of negative $H$ increases further. The origin of these additional modes is discussed later in this paper.

Next, we have investigated the detailed magnetic field dispersion and nature of the observed SW modes. The magnetic field dispersion of SW frequency is plotted in Figure 4a-b for c-ASI and d-ASI, respectively. The experimental BLS spectra taken at varying $H$ values are presented as yellow (high SW intensity) to blue (low SW intensity) heatmaps with numerical simulation results from OOMMF superimposed as filled symbols. The SW modes exhibit distinct and non-monotonic dispersion across the field range investigated.

In Figure 4a, the intensity of the lowest frequency mode M1 is much lower than the other two modes. The frequency splitting between M2 and M3 decreases with the reduction of field from

+1.4 kOe reaching a minimum separation of ~2 GHz between $H$ = -200 Oe to -400 Oe, beyond which it increases again. The frequencies of all modes experience a minimum at $H$ = -400 Oe, which is close to the mid-point of reversal as shown by the black arrow in Figure 2a. For $H <$ -400 Oe, the mode frequencies experience a sharp upwards frequency transition with an increase of ~3 GHz observed in M3. The relative intensities of M2 and M3 are also exchanged, with M2 the dominant mode for $H <$ -400 Oe. As the magnitude of the negative field further increases the intensities of the modes tend towards equal. Figure 4b shows the field dependence of SW frequencies for d-ASI where again three distinct modes are observed. Unlike for c-ASI, the BLS intensity of Mode M2 is highest for d-ASI at positive fields. Here, the lowest frequency mode M1 appears very close to M2. Here also, the frequency splitting between M2 and M3 decreases with decreasing field from +1.4 kOe and they merge together between $H$ = -200 Oe and -400 Oe. Moreover, a crucial distinction between the d-ASI and c-ASI response exists most likely due to the dipole-exchange interaction between nanobars in the c-ASI vertices. After magnetization reversal is complete at -600 Oe, modes M2 and M3 split again with the frequency gap between them increasing with negative field amplitude. The distinct and characteristic SW field dispersion in connected and disconnected kagome ASI demonstrate a high degree of tunability, inviting integration with future spintronic and magnonic devices for microwave filtering applications where the forbidden and allowed frequency regimes may be finely manipulated *via* bias field.

**Numerical Simulations**

To aid interpretation of the experimental findings, the field-dependent static spin configurations and spatial power and phase profiles have been simulated using Object Oriented Micromagnetic Framework (OOMMF) with additional SW analysis performed using in-house developed code.[26] Figure 5a-e show the power profiles of M1-, M1, M1+, M2 and M3 for c-ASI at $H$ = -400 Oe. Here, M1- and M1+ refers to the power maps of frequency slices adjacent to the peak-centre frequency M1. The corresponding phase profiles are shown in the insets at the left corner of each image. We define different quantization numbers $n$ and $m$ for SW modes in backward volume (BV) geometry or in DE geometry, respectively. The mode M1 (Figure 5b) is the edge mode (EM) at the vertex junction of the nanobars. Here strong mode-localization power is concentrated where the two diagonal bars join the horizontal bar. The diagonal nanobars show DE like behaviour with $m = 3$. The similar trend is seen for M1- and M1+ which infers almost no change in the mode characteristics, showing stability of observed mode behaviour over frequency perturbations. Mode M2 shows a BV character in the horizontal nanobar with $n = 9$ and DE character in the diagonal nanobars with $m = 9$. M3 also shows similar character with $n = 11$ and $m = 7$. Modes M2 and M3 exhibit relatively uniform spatial power distributions over the array in comparison with M1. As inter-bar dipole-exchange interaction is strong and low-loss, modes initially excited in one bar subset (horizontal or diagonal) transfer power through vertices and throughout the other subset. This does not occur in vertex-localized mode M1 as while the frequencies of M2 and M3 are close enough to couple between horizontal and diagonal subsets, M1 is excited at substantially lower frequency and hence does not efficiently couple to bar-localized modes M2 and M3. At $H = -550$ Oe, *i.e.* in the transition regime, the number of SW modes have increased to five. Here M1 becomes a BV-like mode with $n = 3$ in the horizontal nanobars, and DE-like mode with $m = 5$. M2 has $n = 9$, $m = 7$ with higher power than for $H = -400$ Oe. M3 has $n = 11$, $m = 7$, similar to M3 for $H = -400$ Oe. M4 has $n = 11$, $m = 9$. M5 has mixed BV-DE character in the individual nanobars and it is difficult to resolve the mode numbers from the phase profile. It is also noteworthy to mention here that some spatial mode localization is evident in the reversed horizontal nanobars (Figure 5i) whereas for diagonal bars, it does not appear as shown in Figure 5j. Therefore, in the case of c-ASI for disordered microstates, mode localization is exhibited preferentially for modes that are oriented along the direction of the applied bias magnetic field, while for bars at an angle to the bias field spatial-dispersion of mode power is observed. As we further increase $H$ to -800 Oe where a complete magnetization reversal occurred to a reverse onion mode, the mode characters are very different from those at $H = -400$ Oe. Here, M1 has $n = 2$, $m = 3$, while M2 has $n = 2$, $m = 5$. Although the phase contrast in the horizontal nanobars on both sides of the nodal plane is not high (far less than $\pi$) and the power is also small, it can still be considered as a BV-like mode with $n = 2$. M3 has $n = 9$ and $m = 9$, similar to M2 at $H = -400$ Oe. This drastic variation in the mode quantization number with the magnetic field leads to the observed abrupt change in the frequencies of the SW modes. Figure 5p-r show the static spin configuration for c-ASI at three bias magnetic fields corresponding to just below (-400 Oe), at (-550 Oe) and above (-800 Oe). Figure 5p shows that the spins in the horizontal nanobars of c-ASI continue to align along the +x direction even at $H = -400$ Oe, while the spins in the diagonal nanobars rotate in the clockwise (anticlockwise) direction in the upper (lower) half of the hexagon forming a forward onion-like state. At $H = -550$ Oe, the spins in certain nanobars start to reverse their orientation in the direction of $H$ forming vortex-like configuration in some of the hexagons in the lower half of the array (shown by ticks ($\sqrt{}$) in Figure 5q). However, in some of the hexagons it forms neither a vortex nor an onion state as shown by asterisks (*) in Figure 5q. When $H$ is further increased to -800 Oe, full reversal of spins in all the nanobars occurs (Figure 5r) forming reverse onion-like states in all hexagons. The corresponding spatial maps of the x-component of demagnetizing field ($H_d$) are presented in the Supporting Information (see Supporting Information, Figure S2). As expected, the direction of $H_d$ lies in the direction opposite to the applied field $H = -400$ Oe. At $H = -550$ Oe, the direction of $H_d$ points towards $H$ only for the reversed nanobars at that field. As we move further to $H = -800$ Oe, the direction of $H_d$ lies completely along $H$. For the clarity of the argument, the simulated SW spectra for c-ASI at various applied bias magnetic field in the negative regime are shown in Supporting Information, Figure S3a.

Next, we investigate the SW mode profiles and static spin configuration for d-ASI at three different applied bias fields, $H$ = -400, -600 and -800 Oe as shown in Figure 6. The simulated power maps are shown together with the corresponding phase profiles in the inset of each image. Here, at $H$ = -400 Oe, M1 corresponds to $n = 5$, $m = 5$, M2 has $n = 3$, $m = 7$, and M3 shows uniform mode (UM), $m = 9$. At $H$ = -600 Oe, M1 has $n = 5$, $m = 5$, M2 has $n = 7$, $m = 7$, M3 has $n = 3$, $m = 7$, M4 has $n = 8$, $m = 9$, M5 has UM, $m = 9$. Hence, M2 and M4 are two additional modes generated at transition, while three other modes retain their characters with some modulation of power and phase distribution. This key observation is also evident in the experimental BLS spectra taken in the transition regime where the partial reversal of nanobars occurred for both c-ASI and d-ASI as shown in Figure 3e-f. For d-ASI, spatial mode localization is clearly evident in all modes. Strong correspondence of modes M1, M2 and M3 with the bar end, diagonal and horizontal bars is observed. Disordered microstates map directly onto the spatial mode power distribution, with M2 and M3 localized in unreversed horizontal and diagonal bars respectively and M4 and M5 corresponding to reversed diagonal and horizontal bars (Figure 6g-j). This starkly contrasts with the spatially dispersed c-ASI modes observed in Figure 5 and occurs due to the relatively weak and inefficient dipolar coupling being unable to transfer considerable spin-wave power outside of bars. This has distinct implications for magnonic system design, showing that for modes to be well spatially localized they must be well-separated in frequency from modes in adjacent elements (as in vertex/bar end localized mode M1) or disconnected from neighbouring elements to preclude mode spreading *via* exchange coupling. At $H$ = -800 Oe, the mode characters have become drastically different. Here M1 corresponds to $n = 4$, $m = 5$ (although M1- and M1+ show $n = 3$, $m = 5$), M2 has $n = 7$, $m = 9$ and M3 has UM, $m = 11$. Again, this drastic change in mode character has led to the observed transition in frequency which show quite different behaviour from $H$ = -400 Oe for d-ASI, while for c-ASI, only one mode gets modified and the other two retain their identity. The characteristic features of these SW spectra as observed from BLS experiment match qualitatively well with that obtained using micromagnetic simulation. The additional modes appear most likely due to the local spin configuration during magnetization reversal as evidenced *via* simulating static spin configuration. As shown in Figure 6p-r, static spin configuration for d-ASI follows a similar trend as c-ASI. However, only one vortex state (√) and three asterisk states (*) are formed in this case (Figure 6q). In addition, the x-component of demagnetizing field for d-ASI is shown in Supporting Information, Figure S2. The direction of $H_d$ in the horizontal nanobars is opposite to that observed in c-ASI. The partial reversal occurred at $H$ = -600 Oe causes local reversal in $H_d$. The striking difference in the frequency of the various SW modes between c-ASI and d-ASI are related with this change in the spin configuration and $H_d$. Moreover, the local modification of the internal field due to the variation of $H_d$ leads to appreciable change in the SW frequencies associated with the reversed nanobars between c-ASI and d-ASI. The simulated SW spectra for d-ASI are shown in Supporting Information, Figure S3b.

In addition, to validate the simulation, a comparative study of the simulation with/without 2D-periodic boundary condition (PBC) performed on a unit cell and large array of ASI for both c-ASI and d-ASI by taking the ideal dimensions of the nanobars is presented in Figure 7. The Simulated SW spectra at a positive saturation field (at $H$ = +1400 Oe) infers the existence of three SW modes in all the cases and the peak frequencies do not vary much as shown in Supporting Information, Figure S4. To investigate the nature of the SW modes, we have simulated the power and phase maps of various SW modes observed at $H$ = +1400 Oe. In Figure 7a-b, the power and phase profiles of all three modes (for c-ASI) simulated for a unit cell with 2D-PBC and large array without applying 2D-PBC are shown, respectively. The characters of the modes are edge mode (EM), BV-like mode (quantization number, *n*) in the horizontal nanobars and DE-like modes (quantization number, *m)* in the tilted nanobars. Here, we see spreading of SW power between bars as opposed to d-ASI (Figure 7c-d). It is most likely due to the weak localization of modes in c-ASI. The SWs are transmitted between bars in c-ASI much more readily due to dipole-exchange mediation at vertices rather than dipolar coupling in d-ASI. In order to emphasize the invariance of the mode characteristics, the SW Mode profiles for the frequency slices adjacent to the central peak is shown in Supporting Information, Figure S5.

One question may arise if these ASIs can show some advances over conventional two-dimensional magnonic crystals. In usual magnonic crystals, SWs are functionalized for applications in on-chip logic and communication devices and/or components by engineering their magnon band structure. The use of ASI as functional magnonic crystals depends upon their reconfigurability of magnetic microstates and the ensuing SW properties.[27] The studied ASI structures exhibit rich SW mode spectra, including mode merging, mode jumping, creation and annihilation of modes and changes in relative mode power with subtle changes in bias magnetic field. In this regard, the c-ASI and d-ASI show distinct differences in quantitative values of the magnetic fields for occurrence in the above features. Besides, in d-ASI the SW modes merge to give rise to a dominant single mode in the lower range of magnetic field on both sides of $H = 0$ as opposed to the c-ASI. Such a reconfiguration of SW spectra stem from the differences in their magnetic microstates. Moreover, the connected and disconnected kagome ASI systems can potentially give rise to a huge variety of magnetic microstates which can be globally or locally controlled simply by a magnetic field[28] or an MFM tip, the so-called topological defect driven magnetic writing[12] as shown before. The d-ASI structure with curved edges can offer additional degree of freedom for magnetization curling at the vertices for self-energy minimization and inter-island interaction.[29] This can lead to very different SW spectra for different edge curling of magnetization at the vertices. The shape anisotropies of the nanobars can be conveniently controlled to create varying coercive fields and large range of magnetic microstates.[24] The c-ASI structure, on the other hand, can be efficiently used for continuous SW nanochannels[30-32] which can be easily reconfigured by rotating the bias magnetic field in the ASI plane.

## CONCLUSIONS

In conclusion, we have experimentally and numerically studied the magnetization dynamics of connected (c-ASI) and disconnected (d-ASI) kagome ASI nanostructures made of Py nanobars using BLS and micromagnetic simulation. The MFM images reveal that the magnetic microstates of kagome ASI obey the spin ice rule and a good agreement is found with the simulated images using micromagnetic simulation. The magnetic hysteresis loop measured by MOKE exhibits two-step reversal process for c-ASI, but a single step reversal for d-ASI. Bias field dependent SW spectra measured by BLS reveal distinct features both in SW frequency ($f$) as well as BLS intensity for c-ASI and d-ASI. Both c-ASI and d-ASI exhibit frequency minimum, occurring at negative bias magnetic fields, when ramped down from a positive saturation field. A sharp jump in the mode frequencies was observed beyond the minima, linked to bar reversal *via* the switching field observed from the MOKE loops and the corresponding mixed magnetization state of nanobars aligned parallel and anti-parallel to the magnetic bias field.

Furthermore, we elucidate striking differences in spatial mode-localization between dipole-exchange-mediated c-ASI and dipolar-coupled d-ASI. This yields useful magnonic design rules for artificial spin systems and the broader family of magnonic crystals: if tight spatial mode-localization is a desired system feature, one must be careful to ensure that if elements are continuously-connected as in c-ASI, modes in adjacent elements must be well-separated in frequency (as between M1 to M2 and M3) else strong dipole-exchange coupling will efficiently transfer power between modes in adjacent element, giving rise to the spatially-dispersed mode profiles in M2 and M3 for c-ASI. This may be mediated by preparing disordered microstates and rotating the applied field to lie along one subset of elements (Figure 5i-j) or by disconnecting elements at the vertex, such that weak dipolar coupling is unable to transfer substantial spin-wave power between adjacent elements. One may harness this effect by continuously-connecting elements of different size and hence differing mode frequency to lithographically-define tailored spatial mode-power profiles,[8] enhancing the degree of reconfigurable magnonic control. The insight gained during this study allows for direct comparison of dipole-exchange *versus* dipolar mediated interaction on spin-wave responses, with strong design implications for both current and future reconfigurable magnonic systems.

**METHODS**

**Sample fabrication**: Arrays of kagome ASI of area $90 \times 90$ µm$^2$ were fabricated using electron-beam lithography (EBL) and lift-off process. Single layer polymethyl methacrylate (PMMA) resist (950K) was used on Si/SiO$_2$ substrate. After writing and development of the resist pattern using MIBK developer, 25-nm-thick Py film was deposited by thermal evaporation at a base pressure of $9.0 \times 10^{-7}$ Torr. The samples were then capped with 3.7 nm of Al$_2$O$_3$ to protect it from oxidation. For the c-ASI sample, the nanobars have length ($l$) and width ($w$) of 300 nm and 80 nm, respectively. The d-ASI has $l$ and $w$ matching the c-ASI but a 50 nm vertex gap (defined as the distance between nanobar-end and vertex centre) is introduced.

Preliminary characterization was performed using SEM to inspect the ASI geometry and nanofabrication quality. MFM was performed at remanence, confirming single-domain nanobars and system microstates obeying the ice rules. Magnetic hysteresis loops were measured using MOKE.

**Brillouin light scattering:** BLS measurements were performed in the DE geometry using a Sandercock-type six-pass tandem Fabry-Pérot interferometer.[33] Conventional 180° backscattered geometry was used to investigate the field evolution of SWs. In the light scattering process, total momentum is conserved in the plane of the thin film. As a result, the Stokes (anti-Stokes) peaks in BLS spectra correspond to creation (annihilation) of magnons with momentum $k = \frac{4\pi}{\lambda}\sin\theta$, where $\lambda$ is the wavelength of the incident laser beam (532 nm in our case), and $\theta$ refers to the angle of incidence of laser. To get well defined BLS spectra for the larger incidence angles, the spectra were obtained after counting photons for several hours. Free spectral range (FSR) of 30 GHz and a $2^{10}$ multi-channel analyzer were used during the BLS measurement. Frequency resolution is determined by estimating FSR/$2^{10}$ ($\approx 0.03$ GHz) for the Stokes and anti-Stokes peaks of the BLS spectra. A variable bias magnetic field ($H$) was applied during the measurements. Further details can be found in reference.[34]

**Micromagnetic simulation:** To support the experimental results, micromagnetic simulations were performed using OOMMF software.[35] The SEM images of the arrays were mimicked and converted into discretized mesh. It is important to mention here that using real SEM images allowed us to introduce the slight discrepancies in the nanobars dimensions resulting due to the nanofabrication imperfections often termed as quenched disorder. A cell size of $5 \times 5 \times 25$ nm$^3$ dimension was employed, with lateral dimensions below the Py exchange length. Typical material parameters of Py were used in simulation, namely: gyromagnetic ratio, $\gamma = 2.211 \times 10^5$ A$^{-1}$ s$^{-1}$, saturation magnetization, $M_S = 800$ kA/m, the exchange stiffness constant, $A = 1.3 \times 10^{-11}$ J/m and Gilbert damping parameter, $\alpha = 0.008$. To simulate the magnetization dynamics, we initially simulated the magnetic microstates at desired bias field values followed by application of an RF excitation field along the array y-axis. The simulated time-varying magnetization was collected for 4 ns at 10 ps steps. LLG micromagnetic simulator[36] was used for simulation of MFM images, using the same material parameters mentioned above.

## ASSOCIATED CONTENT

**Supporting Information**

The Supporting Information is available free of charge on the ACS Publications website [URL will be inserted by publisher]. It consists of the following.

**(I)** Atomic force microscopy (AFM) images. **(II)** Simulated x-component of demagnetizing field, **(III)** Simulated SW spectra for c-ASI and d-ASI at various applied magnetic fields. **(IV)** Comparison between SW spectra simulated under various conditions. **(V)** Simulated SW mode profiles for the frequency slices adjacent to the central peak.

**Notes**

The authors declare no competing financial interest.

**AUTHOR INFORMATION**

**Corresponding Author**

**\*E-mail:** abarman@bose.res.in

**Author Contributions:** A.B. and W.R.B. planned and supervised the project. A.K.C. and J.C.G. prepared the samples and made SEM, MFM and static MOKE measurements in assistance with K.D.S. and A.V. A.K.C. and A.K.M. performed the BLS measurements in consultation with A.B. A.K.C. performed the numerical simulations in assistance with S.B., A.B., J.C.G. and W.R.B. A.K.C. and A.B. wrote the manuscript in consultation with all authors. J.C.G. and W.R.B. thoroughly edited the manuscript.

**Acknowledgements**

A.B. gratefully acknowledges the financial assistance from Department of Science and Technology (DST), Government of India under grant no. SR/NM/NS-09/2011 and S. N. Bose National Centre for Basic Sciences under project no. SNB/AB/18-19/211. A.K.C. gratefully acknowledges DST, Government of India for INSPIRE fellowship (IF150922) and British Council UK for providing Newton Bhabha fellowship under Newton Bhabha Fund PhD Placement Program (Award No. DST/INSPIRE/NBHF/2018/4). A.K.M. acknowledges S. N.

Bose National Centre for Basic Sciences, India for senior research fellowship. S.B. acknowledges Science and Engineering Research Board (SERB), India for funding (grant no. CRG/2018/002080). W.R.B. gratefully acknowledges financial support by the Leverhulme Trust (RPG-2017-257). A.V. was supported by the EPSRC Centre for Doctoral Training in Advanced Characterization of Materials (Grant No. EP/L015277/1).

**List of Figures:**

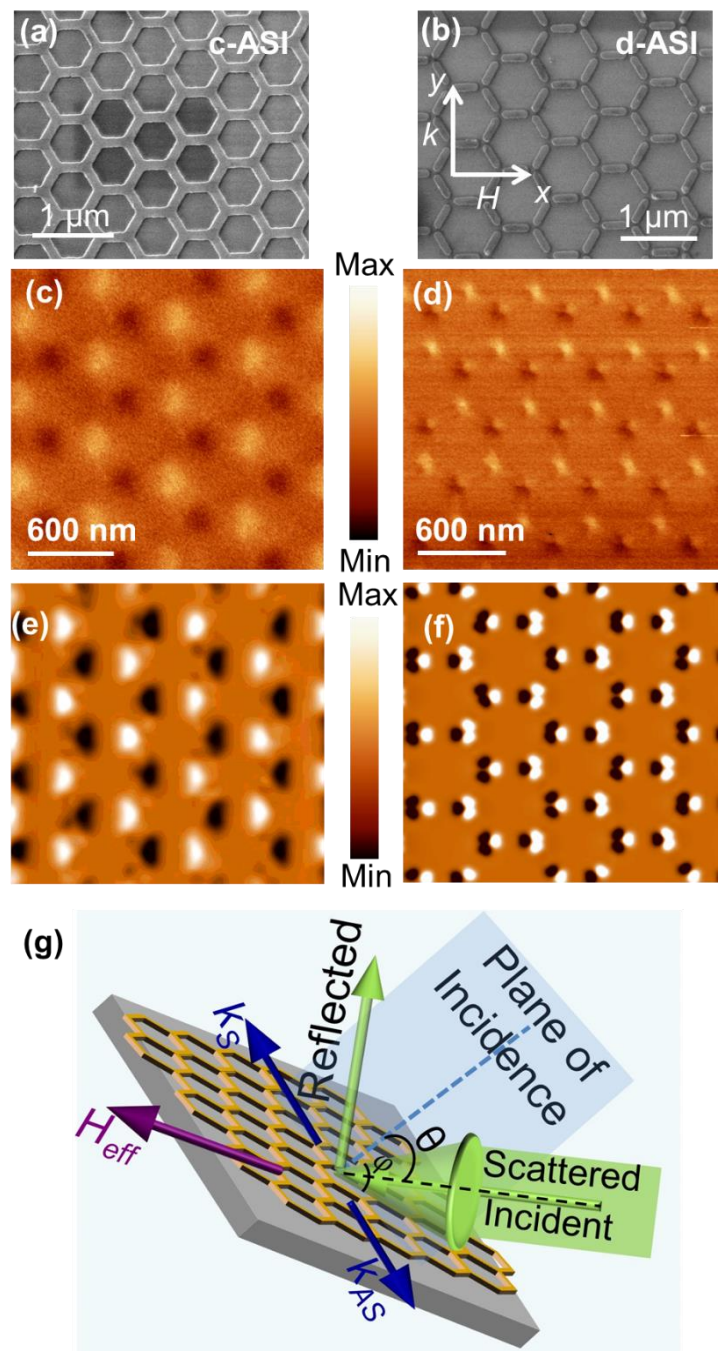

**Figure 1.** SEM images of (a) connected (c-ASI) and (b) disconnected (d-ASI) kagome ASI. Experimental MFM images taken at remanence for (c) c-ASI and (d) d-ASI. Simulated MFM images at remanence for (e) c-ASI and (f) d-ASI. (g) Schematic of the Brillouin light scattering measurement geometry.

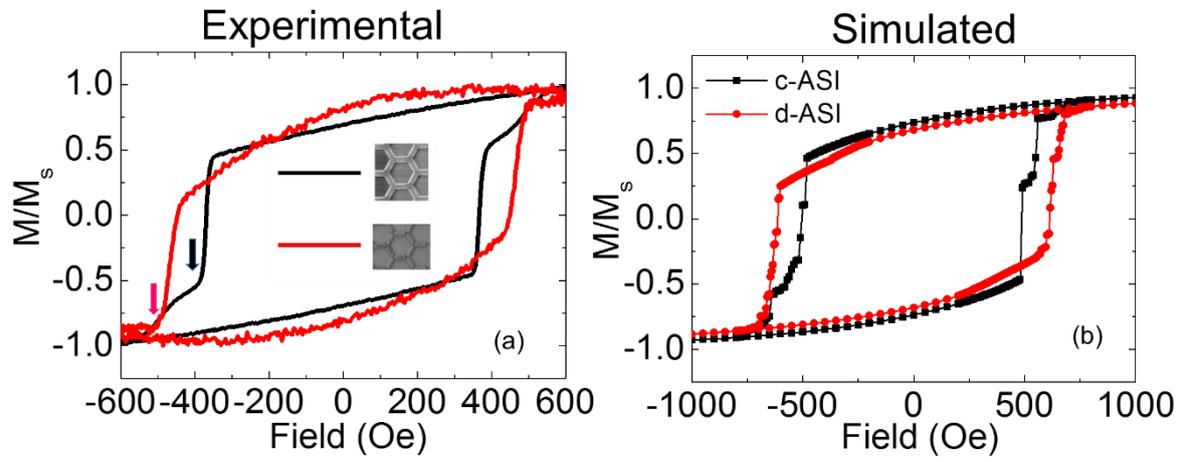

**Figure 2.** (a) MOKE hysteresis loops showing the magnetization reversal for connected (c-ASI) and disconnected (d-ASI) kagome ASI samples. Arrows indicate fields at which BLS data was collected in Figure 3e-f. (b) Simulated hysteresis loop for both the samples.

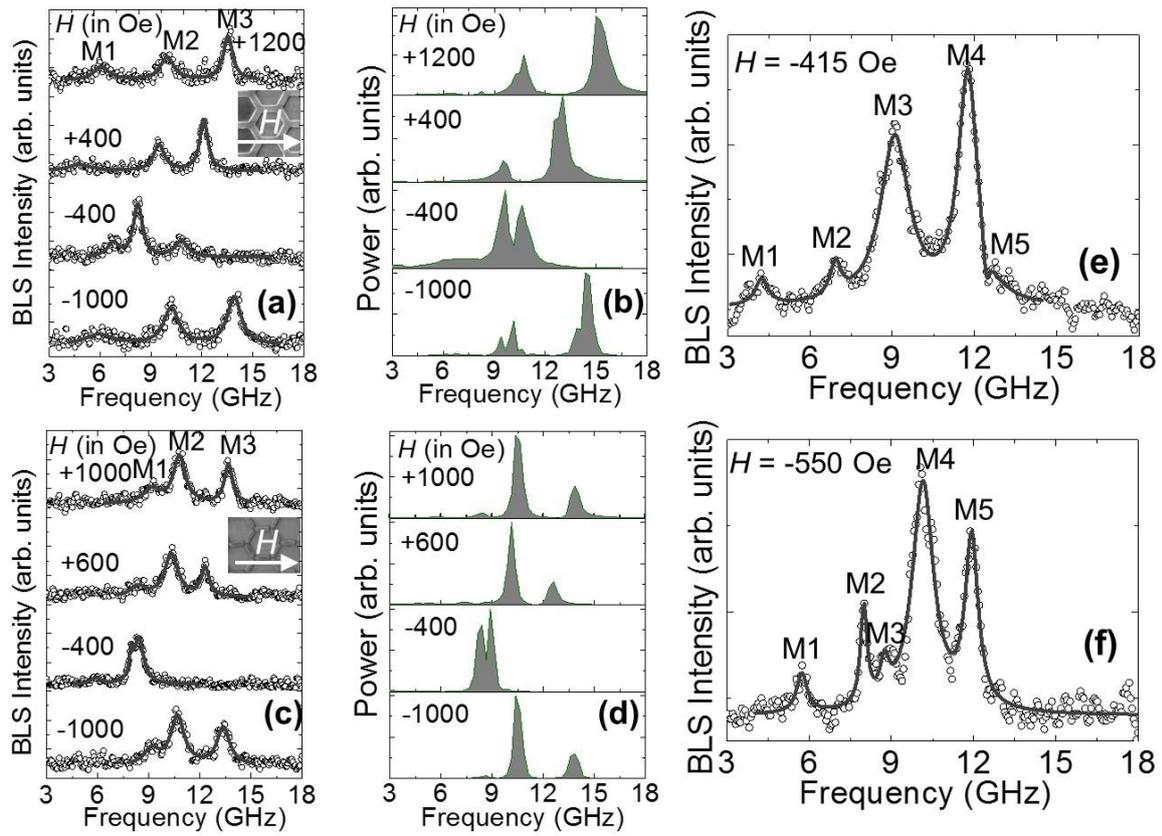

**Figure 3.** Representative BLS spectra measured at wave vector $k \approx 0$ for different magnetic fields from (a) c-ASI and (c) d-ASI. Each spectrum corresponds to a different magnetic field value as indicated next to the spectrum. The solid grey curves represent Lorentzian multipeak fits. The SW modes M1, M2, and M3 are marked in ascending peak frequency. Simulated SW spectra are shown for c-ASI (b) and d-ASI (d). BLS spectra taken halfway through magnetization reversal are shown for c-ASI (e) and d-ASI (f) with additional modes present relative to the saturated state.

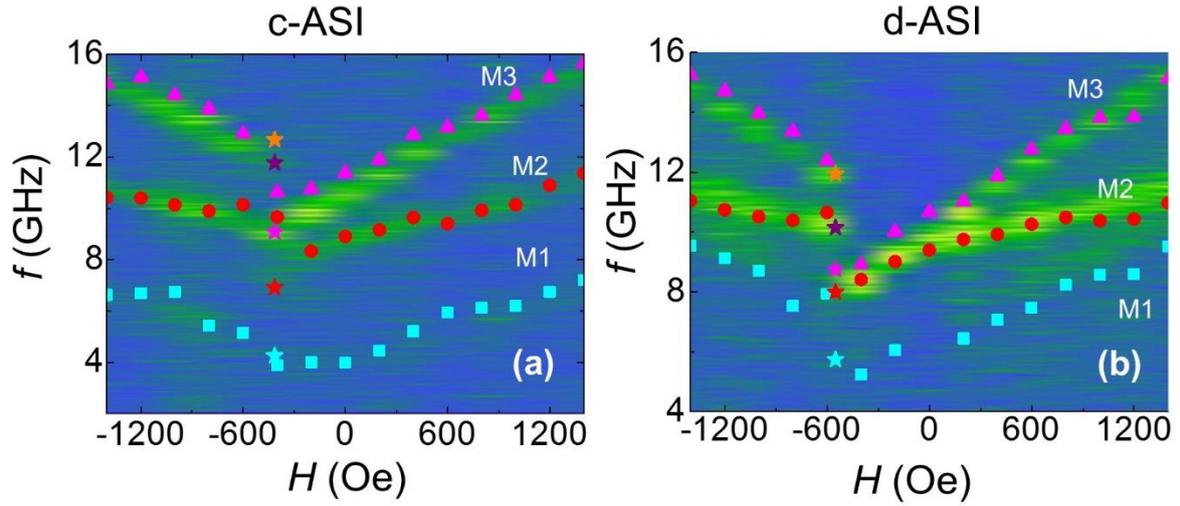

**Figure 4.** SW mode frequencies of (a) c-ASI and (b) d-ASI as a function of applied bias magnetic field along the horizontal bar axis. The experimental bias field dependent SW frequencies are shown as heat maps with yellow region denoting high SW intensity. Simulated SW mode frequencies are shown by filled symbols. The different SW branches are denoted by M1, M2 and M3. The experimental data taken halfway through magnetization reversal (corresponding to the SW spectra shown in Figure 3e-f) are denoted by star symbols for both c-ASI and d-ASI.

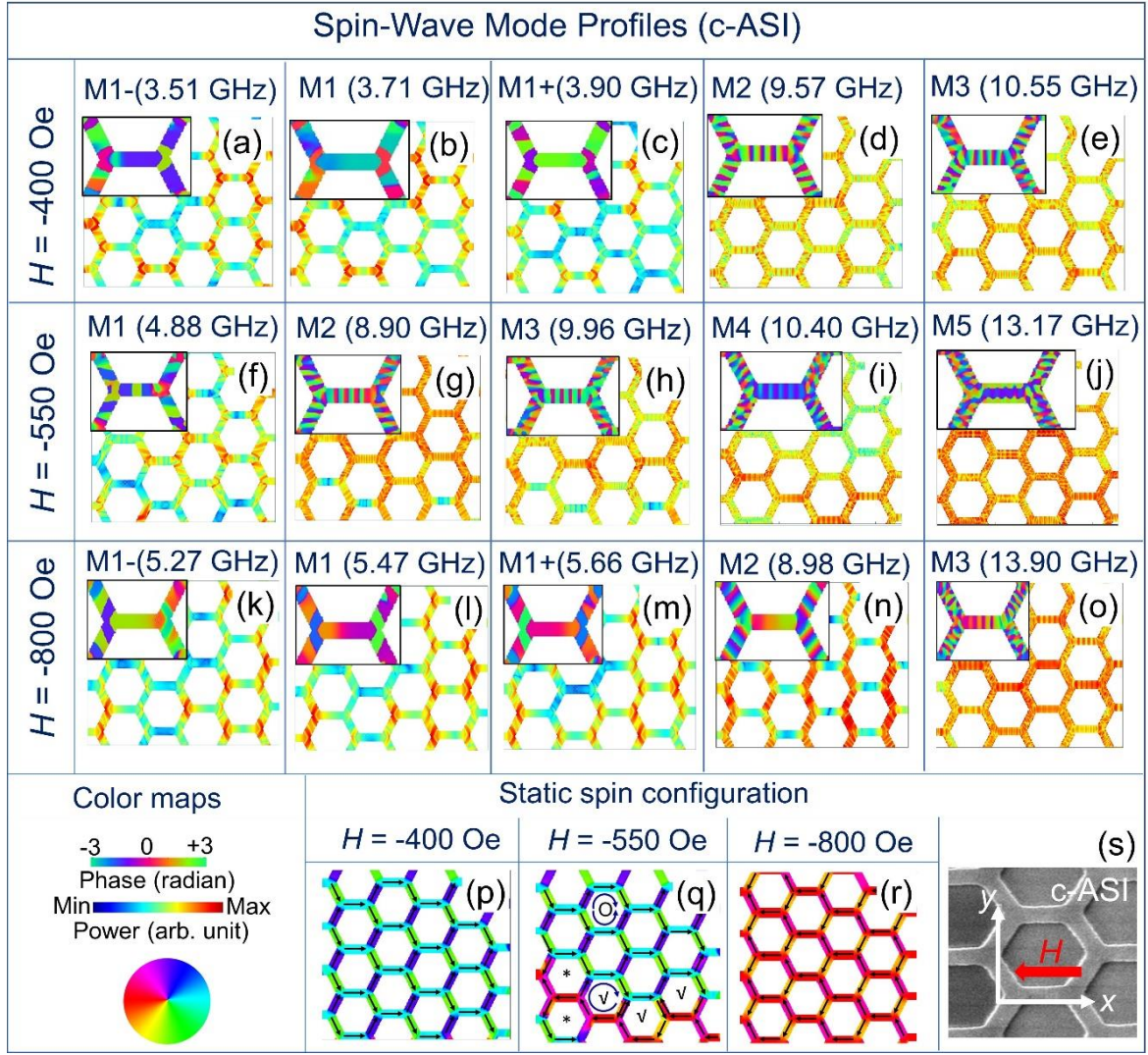

**Figure 5:** Simulated SW power profiles at $H$ = -400 Oe (a-e), -550 Oe (f-j) and -800 Oe (k-o) for c-ASI. M1- and M1+ refers to the power maps of frequency slices adjacent to the peak-centre frequency M1. Phase profiles are shown in the inset at the left corner of each power profile. Figure (p-r) shows the simulated static spin configuration. Other than vertex-localized behaviour observed in mode M1, a low-degree of mode spatial confinement is observed in c-ASI. Modes are seen to spread throughout the connected lattice due to strong dipole-exchange coupling at vertices, with near equal power through the array observed in (d, j, o). The vortex like configuration in some of the hexagons are marked with tick (√) symbol. The star (*) marks refer to the hexagons which forms neither vortex nor an onion state. The flow of the magnetization is shown by curved arrows. The color maps are given at the bottom left corner. (s) The SEM image of the c-ASI sample along with the measurement geometry is shown at the bottom right corner.

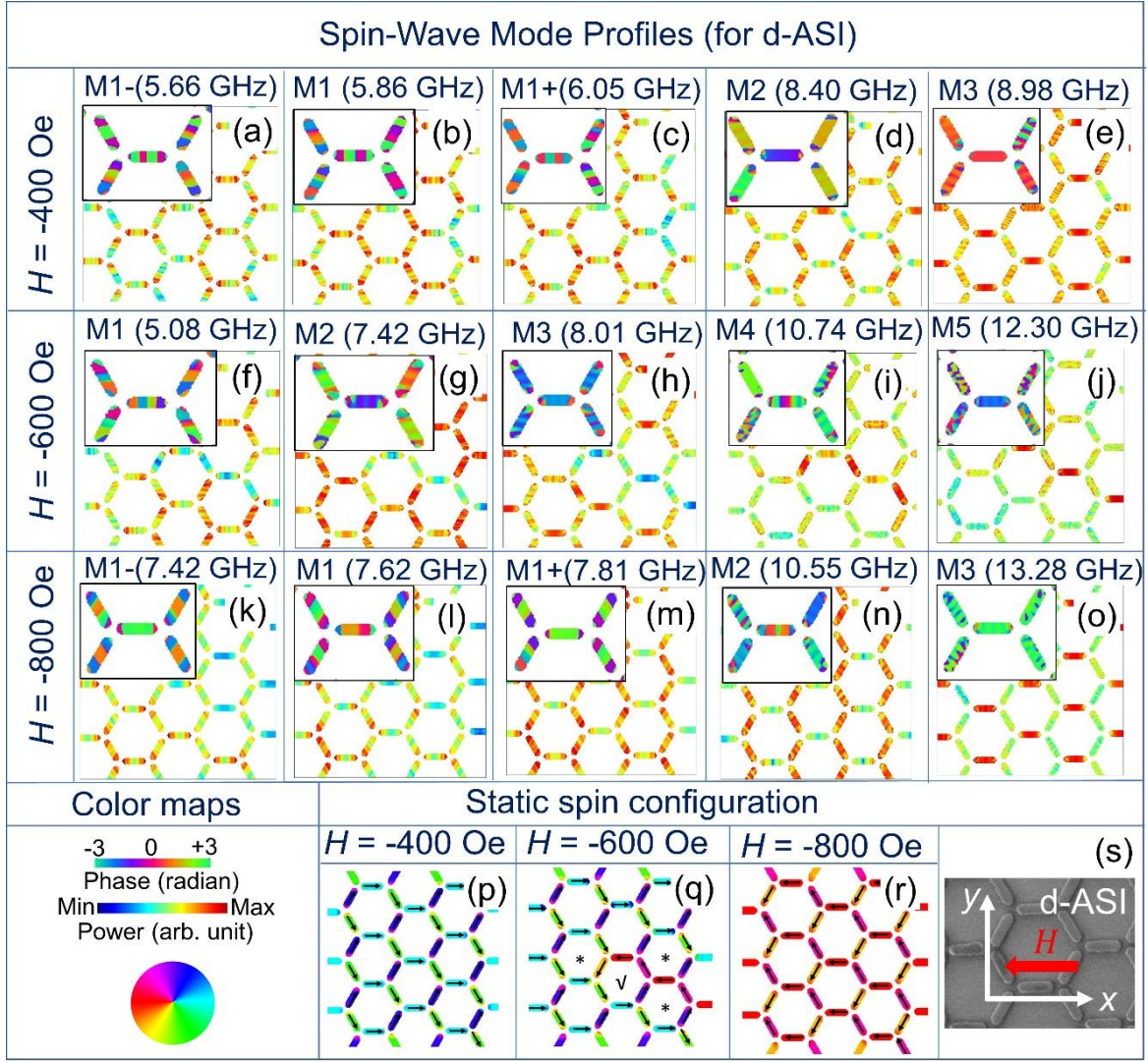

**Figure 6**: Simulated power maps for various precessional modes at $H$ = -400 Oe (a-e), -600 Oe (f-j) and -800 Oe (k-o) for d-ASI. M1- and M1+ refer to the power maps of frequency slices adjacent to the peak-centre frequency M1. Phase profiles are shown inside the rectangular box at the left corner of each image. (p-r) shows the simulated static spin configuration. Spin-wave mode power exhibits a high-degree of spatial localization, with mode-power at each frequency showing a strong correspondence to the static spin configurations in (p-r). The vortex like configuration in some of the hexagons are marked with tick (√) symbol. The asterisk (*) marks refer to the hexagons which forms neither vortex nor an onion state. The flow of the magnetization is shown by curved arrows. The colour maps are given at the bottom left corner. (s) The SEM image of the d-ASI sample along with the measurement geometry is shown at the bottom right corner.

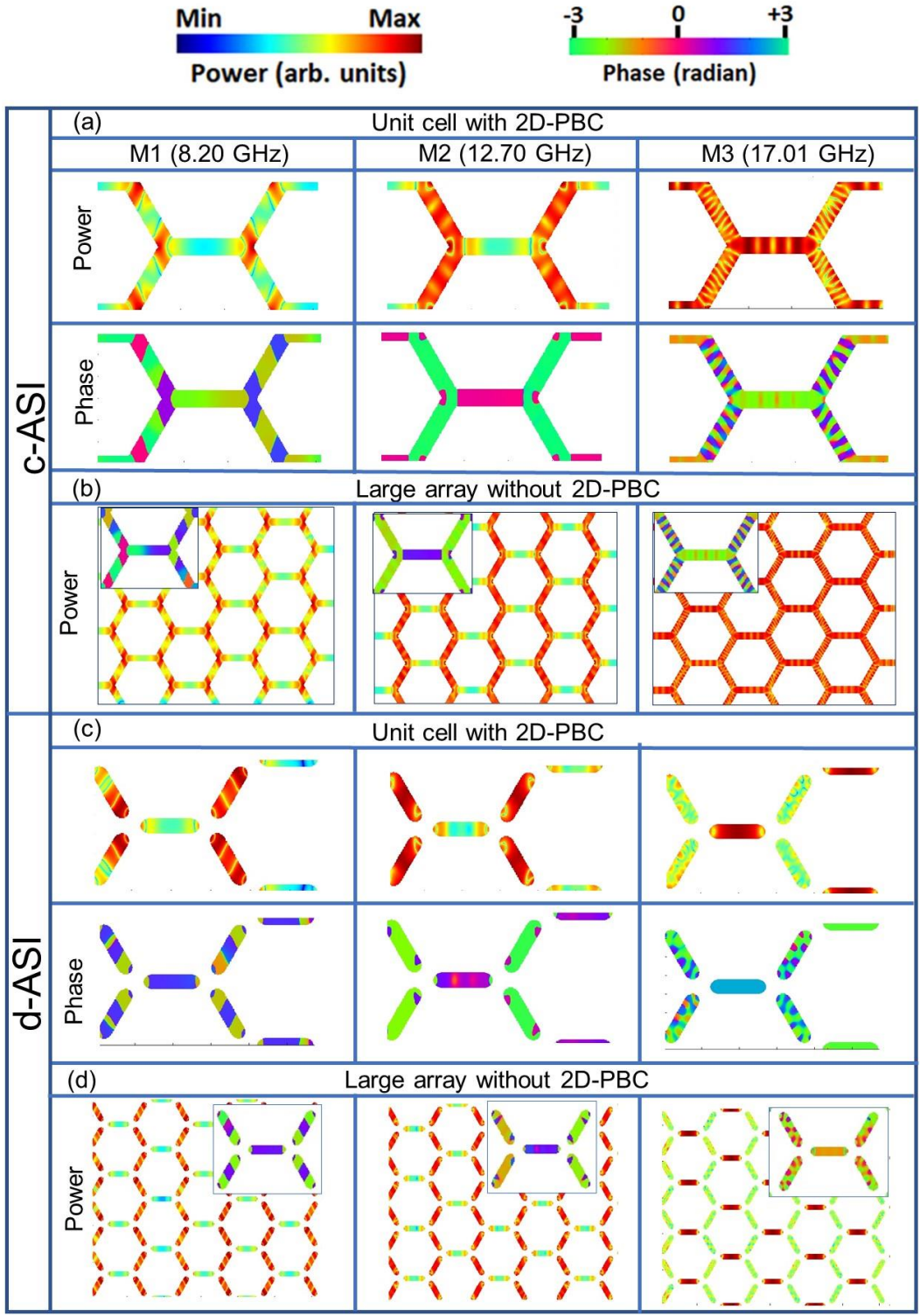

**Figure 7:** Simulated power maps for various precessional modes in c-ASI and d-ASI at $H$ = 1400 Oe for unit cell with 2D-PBC and large array without 2D-PBC. Phase profiles are shown inside the rectangular box at the left corner of each image. The color maps are shown on the top.

**TABLE OF CONTENT**

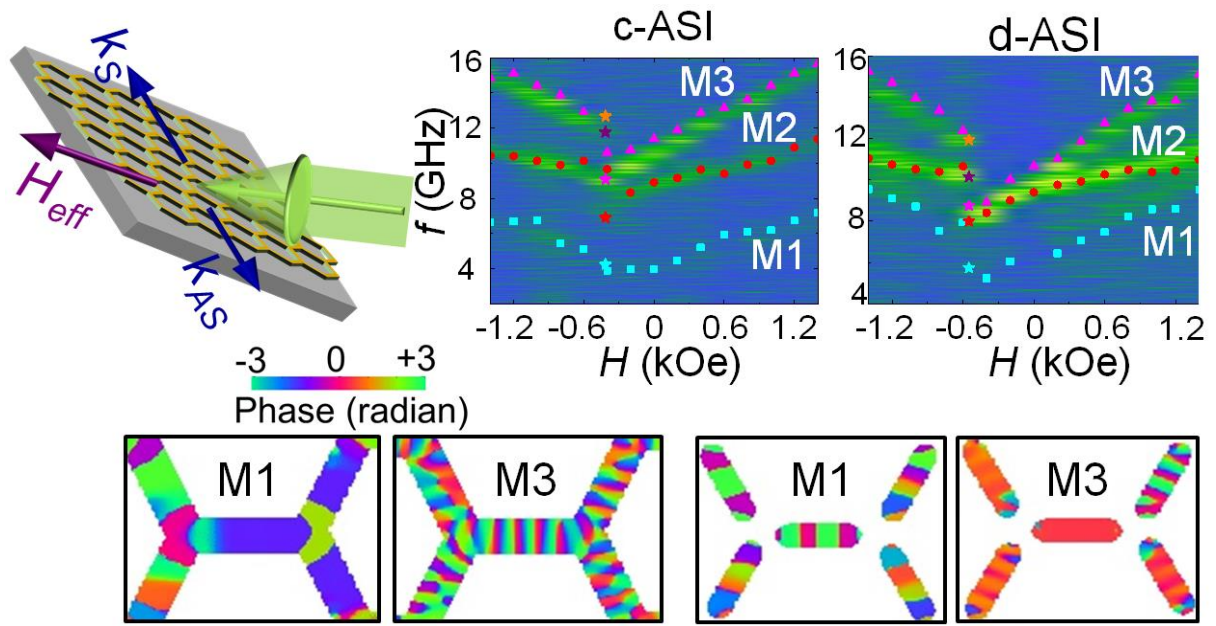

Spin-wave dynamics in kagome artificial spin ice nanostructure

28